\documentstyle[12pt,moriond,epsfig]{article}

\def\be{\begin{equation}}
\def\ee{\end{equation}}
\def\bea{\begin{eqnarray}}
\def\eea{\end{eqnarray}}

\begin{document}
\vspace*{4cm}
\title{ESTIMATION OF PARAMETERS OF GRAVITATIONAL WAVES FROM PULSARS}

\author{ A. KR\'OLAK }

\address{ Max Planck Institute for Gravitational Physics\\ The Albert
Einstein Institute\\ Potsdam, Germany \footnote{On leave of absence 
from Institute of Mathematics, Polish Academy of Sciences, Warsaw,
Poland.}} 

\maketitle\abstracts{The problem of search for nearly periodic 
gravitational wave sources in the data from laser interferometric
detectors is discussed using a simple model of the signal.
Accuracies of estimation of the parameters and computational
requirements to do the search are assessed.}

\section{Introduction}

Pulsars are one of the primary sources of gravitational-waves
that can be observed by detectors that are currently under construction
\cite{Liv,Sch,Jon}. The data analysis involved to
do the search for such sources implies a very heavy 
computational cost \cite{Sch}.
Here we analyse this problem using a simple model of the 
gravitational-wave signal from a pulsar. 
A detailed summary of the current understanding of
the gravitational-wave
pulsar phenomenology and an analysis of an efficient data analysis
technique based on a more accurate model of the signal has recently 
been presented \cite{BCCS}.

\section{A simple model of the gravitational-wave signal from a pulsar}

The frequency of the gravitational-wave signal from a pulsar will follow
its rotational frequency and therefore the signal is expected to be
almost monochromatic.
However the amplitude of the signal will be very small and to extract
it from the noise we may require to integrate the data for several months.
Consequently the modulation of the signal due to the motion of the
detector relative to the solar system barycenter and even very small
change of the frequency of the pulsar will need to be taken into account.

Here we consider a simple model of the signal where we take into account
only the modulation of the signal due to the motion of the Earth
around the Sun and we approximate the change of frequency during
the observation time by a Taylor series \cite{Cur}. 
Let $R_{\odot}$ be 1 astronomical unit (AU), $\Omega = 2 \pi/1\mbox{year}$
and let the position of the pulsar on the sky be $(\theta,\phi)$
in the coordinate system based on the ecliptic (i.e., $\theta = \pi/2$ is
Earth-Sun plane, and $\phi =0$ is position of Earth at $t = 0$).
Let $\omega(t)$ be the angular gravitational-wave frequency from the
pulsar.
We approximate the phase $\int^t \omega(t')\,dt'$ of the signal
by a power series $\omega_1 t + \omega_2 t^2 + \omega_3 t^3 + 
\omega_4 t^4 + \phi_o$, where $f_1=\omega_1/2\pi$ is the frequency of the
pulsar at an arbitrarily chosen instant of time $t_o$ and $\omega_{1+s}$
is the $s$th spin-down parameter proportional to the $s$th derivative
of the frequency at time $t_o$ and $\phi_o$ is a constant phase.
The number of terms needed in the expansion depends
on the observation time and the expected values of the frequency
derivatives.
 
We can introduce the following estimates for the spin-down parameters:
\be
|\omega_{1+s}| \simeq \frac{\omega_1}{\tau^s}x^s,
\ee
where $\tau$ is the age of the pulsar and $x_s \leq 1$. 
One can expect that for young pulsars
$x^s$ are of the order of $1$ and less than $1$ for old pulsars. 
Thus we have the following model of the gravitational-wave signal:
\be
h(t) = h_o \sin [\omega_1 t + \omega_2 t^2 + \omega_3 t^3 +  
\omega_4 t^4 +
\omega_1 t_{\odot}\sin\theta \cos(\Omega t - \phi) + \phi_o],
\label{sig}
\ee
where $h_o$ is the constant amplitude and $t_{\odot}=R_{\odot}/c$.
The amplitude $h_o$ is estimated as \cite{Kip}
\be
h_o = 7.7 \times 10^{-24} (\frac{I_{\bar{z}\bar{z}}}{10^{45}\mbox{g
cm}^2})
(\frac{1\mbox{kpc}}{r})(\frac{f_1}{1\mbox{kHz}})^2(\frac{\delta}{10^{-5}}),
\ee
where $I_{\bar{z}\bar{z}}$ is the moment of inertia of the pulsar about
its rotation axis, $r$ is the distance, $f_1$ is the gravitational wave
frequency and $\delta$ is the ellipticity of the pulsar.
The ellipticity of $10^{-5}$ is an estimate corresponding to the maximum
strain that the neutron star crust may support.
In a realistic model a number of other corrections
will need to be taken into account \cite{BCCS}.

\section{Data analysis technique}

The signal given by Eq.(\ref{sig}) will be buried in the noise of the 
detector. Thus we are faced with the problem of detecting the signal
and estimating its parameters.
A standard method is the method of {\em maximum likelihood detection} 
which consists of maximizing the likelihood function $\Lambda$ with
respect to the parameters of the signal. If the maximum of $\Lambda$
exceeds a certain threshold calculated from the false alarm probability
that we can afford we say that the signal is detected. The values of the
parameters that maximize $\Lambda$ are said to be {\em maximum likelihood
estimators} of the parameters of the signal.
The magnitude of the maximum of $\Lambda$ determines the probability of
detection of the signal.
We assume that the noise $n$ in the detector is an additive, stationary,
Gaussian, zero-mean random process. Then the log likelihood function has
the form
\be
\log\Lambda= (x|h) -\frac{1}{2}(h|h),
\ee
where $x$ are the data and $h$ is the signal and the scalar
product is defined as
\be
(x|h) =
4\Re\int^{\infty}_{0}
\frac{\tilde{x}(f)\tilde{h}^{*}(f)}{S_h(f)}\,df,
\ee
where $\tilde{}$ denotes Fourier transform, * is complex 
conjugation, and $S_h(f)$ is the spectral density of the noise.
We can assume that during the time of observation the signal from
the pulsar is almost monochromatic. 
Hence we can approximate the scalar product as
\be
(x|h) \simeq \frac{2}{S_h(f_1)}\int^{T/2}_{-T/2}x(t)h(t)\,dt,
\ee
where $T$ is the observation time.
We can write our signal as $h = h_o\cos\phi_o h_c + h_o\sin\phi_o h_s$.
Since during the observation time the signal will have many cycles
and frequency will not change appreciably, to a very good approximation
we have
\bea
(h_c|h_s) &=& 0, \label{sc1}\\
(h_c|h_c) &=& (h_s|h_s) = H_o ,
\label{sc2}
\eea
where $H_o$ is a constant $\simeq \frac{T}{S_h(f_1)}$. 
We can find closed analytic expressions for the maximum likelihood
estimators of the amplitude $h_o$ and the phase $\phi_o$ of the signal.
Substituting these expressions into $\log\Lambda$ and using
Eqs.(\ref{sc1}) and (\ref{sc2})
we get the following formula for the reduced likelihood
function ${\cal F}$ which now depends only on the parameters $\omega_i$
and the parameters $\theta, \phi$ determining the position of the source
in the sky:
\be
{\cal F} = \frac{(x|h_c)^2 + (x|h_s)^2}{2 H_o}. 
\ee
For pulsar signal case this last expression can be approximated as
\bea
{\cal F} &\simeq& \frac{2}{S_h(f_1) T}
[(\int^{T/2}_{-T/2}x(t)h_c(t)\,dt)^2 +
(\int^{T/2}_{-T/2}x(t)h_s(t)\,dt)^2] \\
&=& \frac{2}{S_h(f_1) T}|\int^{T/2}_{-T/2}x(t)\exp[-i\Phi_m(t) - i\omega_1 
t]\,dt|^2 = 
\frac{2}{S_h(f_1) T} |\tilde{y}|^2,
\eea
where $\Phi_m(t) = \omega_2 t^2 + \omega_3 t^3 +  \omega_4 t^4 +
\omega_1 t_{\odot}\sin\theta \cos(\Omega t - \phi)$,
$y(t)$ is the data multiplied by $\exp[-i\Phi_m(t)]$ and 
the rectangular window function which is equal to $1$
over the time interval $[-T/2, T/2]$ and zero otherwise.
Tilde denotes the Fourier transform.
The above calculation suggests one way of evaluating
the optimum statistics ${\cal F}$: multiply the data by
$\exp[-i\Phi_m(t)]$ and perform the Fourier transform.
This leads to an efficient algorithm since we can
use the fast Fourier transform. 

The probability of detection of the signal is determined by the 
signal-to-noise ratio $d$ given by $d = (h|h)^{1/2} \simeq 
h_o\sqrt{\frac{T}{S_h(f_1)}}$.
In Table 1 we summarize the numerical values
for signal-to-noise ratios that can be achieved
by laser interferometers currently under construction
by LIGO, VIRGO, GEO600, and TAMA projects.
We choose pulsar with $I_{\bar{z}\bar{z}} = 10^{45} \mbox{g cm}^2$,
$\delta = 10^{-5}$. The gravitational-wave frequency $f_1$ is $215$Hz
and the observation time $T$ is $10^7$s.
INITIAL assumes approximate model for the noise of LIGO and VIRGO
detectors at the beginning of their operation.
ADVANCED assumes their ultimate sensitivity.
\begin{table}
\caption{Signal-to-noise ratios for pulsar signals.\label{tab:exp}}
\vspace{0.4cm}
\begin{center}
\begin{tabular}{|c|c|c|c|c|}
\hline
  & INITIAL & ADVANCED & GEO600 & TAMA \\
\hline
d & 88 & 320 & 14 & 3 \\
\hline
\end{tabular}
\end{center}
\end{table}
The distance to the pulsar is taken to be 1kpc except for TAMA
where it is 0.1kpc.
 
The rms errors of the estimators of the parameters of the signal
are approximately given by the square roots of the diagonal
elements of the inverse of the Fisher information matrix $\Gamma_{ij}$
given by
\be
\Gamma_{ij} =
\left(\frac{\partial{h}}{\partial{\theta_i}}|
\frac{\partial{h'}}{\partial{\theta'_j}}\right)_{|\theta_i=\theta'_j} =
d^2\frac{\partial^2{G}}{\partial{\theta_i}
\partial{\theta'_j}}_{|\theta_i=\theta'_j},
\ee
where $h'$ is the signal in terms of parameters $\theta'_i$ and
in our case
\be
G = \frac{1}{T}\int^{T/2}_{-T/2}\cos[\Phi(t,\theta_i) 
- \Phi(t,\theta'_i)]\,dt.
\label{C}
\ee
$\bf{\Gamma}^{-1}$ is called the covariance matrix and it is denoted
by $\bf{C}$.
Instead of the angles $\theta$ and $\phi$ it is convenient
to introduce the following parameters $a$ and $b$
\bea
a &=& \omega_1 t_{\odot}\sin\theta \cos\phi, \\
b &=& \omega_1 t_{\odot}\sin\theta \sin\phi.
\eea
In this new parametrization the phase of the signal has the form
\be
\Phi(t) = \omega_1 t + \omega_2 t^2 + \omega_3 t^3 +  
\omega_4 t^4 + a\cos(\Omega t) + b\sin(\Omega t) + \phi_o\
\ee
and it is a linear function of the parameters.
As a result the correlation function $G$ depends only on
the difference between the values of the parameters and not
on their absolute values.
Consequently the components of the $\bf{\Gamma}$ matrix are independent
of the values of the parameters.

\section{Numerical values of the rms errors of the estimators
of the parameters of the pulsar signal}

For observation times $T$ less than about 1/3 of a year one
can approximate the components of the covariance matrix
to a very good accuracy by its leading term in the series expansion
in $T$. 
Let $\sigma_{\theta_i}$ be the square root of the
component $C_{\theta_i\theta_i}$ of the covariance matrix.
It is convenient to express the errors in the parameters
$\omega_i$ by the following dimensionless quantities
\bea
\delta^r_{f}&=&\frac{\sigma_{\omega_1}}{\omega_1} \simeq
7.9 \times
10^{-9}(\frac{1/3\mbox{yr}}{T})^5(\frac{1\mbox{kHz}}{f_1})(\frac{10}{d}),\\
\delta^r_{1}&=&\frac{\sigma_{\omega_2}}{\omega_1/\tau} \simeq
1.1 \times 10^{-5}
(\frac{1/3\mbox{yr}}{T})^{6}(\frac{1\mbox{kHz}}{f_1})
(\frac{\tau}{40\mbox{yr}})(\frac{10}{d}),\\
\delta^r_{2}&=&\frac{\sigma_{\omega_3}}{\omega_1/\tau^2} \simeq
8.3 \times 10^{-5}
(\frac{1/3\mbox{yr}}{T})^5(\frac{1\mbox{kHz}}{f_1})(\frac{\tau}{40\mbox{yr}})^2
(\frac{10}{d}),\\
\delta^r_{3}&=&\frac{\sigma_{\omega_4}}{\omega_1/\tau^3} \simeq
6.0 \times 10^{-2}
(\frac{1/3\mbox{yr}}{T})^6(\frac{1\mbox{kHz}}{f_1})
(\frac{\tau}{40\mbox{yr}})^3(\frac{10}{d}).
\label{sp}
\eea
The above equations give lower bounds on the rms errors of the spin down
parameters.
The rms error $d\Omega$ in the position of the source in the sky is given by
\be
d\Omega = \pi \sin\theta\sigma_{\theta}\sigma_{\phi} \simeq
1.3 \times 10^{-6} |\frac{\sin 2\phi}{\cos\theta}|(\frac{1/3\mbox{yr}}{T})^{12}
(\frac{1\mbox{kHz}}{f_1})^2 (\frac{10}{d})^2 \mbox{sr},
\ee
where $\sigma_{\theta}$ and $\sigma_{\phi}$ are rms errors in the
position angles $\theta$ and $\phi$ respectively.

\section{Computational requirements}
To detect the signal and find the estimators of the parameters we need
to find the maximum of the functional ${\cal F}$ with respect
to the parameters.
The computational burden of the search over the parameter $\omega_1$
is minimized because we can take advantage of the speed of the FFT 
algorithm.
The search over the other parameters can be performed by means
of a bank of filters (templates).
The filtered noise $(n|h)$ can be thought of as a multi-dimensional
random process M$(\theta_i)$ with correlation function given by
Eq.(\ref{C}).
In our simple model the correlation function depends only on the
difference between the parameters and the random process is 
a generalization of a stationary random process.
We can generalize the concept of correlation time to such processes,
defining the correlation hyperellipsoid of the process.
The number of independent samples N of such 
an process can be defined as the ratio of the volume of
the parameter space $V$ over the area of the correlation hyperellipsoid.
\be
\mbox{\cal{N}} = \frac{V}{(\pi^{n/2}/\Gamma(n/2+1))\sqrt(detC_{ij})},
\label{NT}
\ee
where $n$ is dimension
of the parameter space and $C_{ij}$ is the covariance matrix.

We use the above formula to estimate the number of independent
filters needed to probe the signal parameter space.

For our case since the phase can be eliminated from the
search and since to estimate the parameter $\omega_1$ we use the FFT
we  insert in the above formula the {\em reduced} covariance 
matrix which is an $n$ by $n$ submatrix of the covariance matrix
corresponding to the $n$ parameters that we search for.

The volume $V$ of the parameter space is given by
\be
V \simeq \pi (\omega_{1max} t_{\odot})^2\omega_{1max}^s
(\tau_{min})^{-s(s+1)/2},
\ee
where $\omega_{1max}$ is the maximum frequency we search for
and $\tau_{min}$ is the minimum spin-down time.

For observation times less than about 1/3 of a year the number of
templates can well be approximated by the leading terms of the Taylor
expansion of Eq.(\ref{NT}).

We obtain the following formulae 
\bea
N_0 &\simeq&  4.7 \times 10^{10}
(\frac{T}{1/3\mbox{yr}})^5(\frac{f_{1max}}{1\mbox{kHz}})^2,\\
N_1 &\simeq&  3.4 \times 10^{16}
(\frac{T}{1/3\mbox{yr}})^{11}(\frac{f_{1max}}{1\mbox{kHz}})^3
(\frac{40\mbox{yr}}{\tau_{min}}),\\
N_2 &\simeq&  3.4 \times 10^{19}
(\frac{T}{1/3\mbox{yr}})^{14}(\frac{f_{1max}}{1\mbox{kHz}})^4
(\frac{40\mbox{yr}}{\tau_{min}})^3,\\
N_3 &\simeq&  5.5 \times 10^{19}
(\frac{T}{1/3\mbox{yr}})^{20}(\frac{f_{1max}}{1\mbox{kHz}})^5
(\frac{40\mbox{yr}}{\tau_{min}})^6.
\eea
The indicies $1, 2, 3$ mean that 1, 2, and 3 spin-down
parameter were included in the calculation.
The exact values are plotted in Figure 1.

From Figure 1 we see that at certain observation times the
curves intersect. The intersection
points give times of observation at which we should include a
next spin-down parameter in the search \cite{BCCS}.
To find the number of templates for a given observation time one 
takes the largest of the numbers $N_i$ given above.

For directed searches for a pulsar of known position in the sky
we obtain
\bea
N_{01} &=&  2.1 \times 10^7
(\frac{T}{1/3\mbox{yr}})^2(\frac{f_1}{1\mbox{kHz}})
(\frac{40\mbox{yr}}{\tau_{min}}),\\
N_{02} &=&  1.1 \times 10^{12}
(\frac{T}{1/3\mbox{yr}})^5(\frac{f_1}{1\mbox{kHz}})^2
(\frac{40\mbox{yr}}{\tau_{min}})^3,\\
N_{03} &=&  1.5 \times 10^{14}
(\frac{T}{1/3\mbox{yr}})^9(\frac{f_1}{1\mbox{kHz}})^3
(\frac{40\mbox{yr}}{\tau_{min}})^6.
\eea
The latter formulae are exact.
The number of floating point operations per second (flops) 
required to perform a search can be obtained by multiplying
the above
formulae by number of operations required to calculate the modulus
of the Fourier transform ($3N(\log N + 1/2)$, where $N = 2 f_{1max} T$, is
the number of points of each FFT) and dividing by the time of
analysis. Assuming that the computation should proceed at
the rate of data acquisition
and for $T=30$ days, $f_{1max}=1$kHz, $\tau_{min}=40$yr
the computing power required is around $4\times 10^4$ Tflops.

\begin{figure}
\epsfig{figure=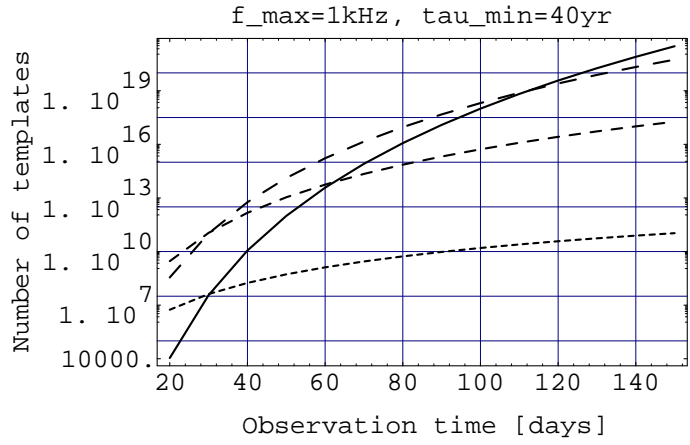}
\caption{Number of templates.\label{fig:radish}}
\end{figure}

\section*{Acknowledgments}
I would like to thank Patrick R. Brady and Bernard F. Schutz
for very helpful discussions
and Piotr Jaranowski for help in numerical work. 
This work was supported in part by Polish Science
Committee grant KBN 2 P303D 021 11.


\section*{References}
\begin{thebibliography}{99}
\bibitem{Liv} J C Livas in {\em Gravitational Wave Data Analysis}, ed. B F
Schutz (Kluwer, Dordrecht, 1989).
\bibitem{Sch} B F Schutz in {\em The Detection of Gravitational Waves},
ed. D.G. Blair (Cambridge University Press, Cambridge, 1991).
\bibitem{Jon} G S Jones, Ph.D. thesis, University of Wales, 1995.
\bibitem{Cur} C Cutler, Unpublished notes, 1996.
\bibitem{Kip} K S Thorne in {\em Three Hundred Years of Gravitation},
ed. S W Hawking and W Israel 
(Cambridge University Press, Cambridge, 1987).
\bibitem{BCCS} P R Brady T Creighton C Cutler and B F Schutz,
Searching for periodic sources with LIGO, submitted
for publication, gr-qc 9702050.
\end{thebibliography}
\end{document}